\documentclass[final,twocolumn]{elsarticle}

\usepackage{amssymb}

\usepackage{amsmath}

\usepackage{lineno}
\usepackage{xcolor}
\usepackage{graphicx}
\usepackage[hyphens]{url}
\usepackage[linesnumbered,ruled,vlined]{algorithm2e}
\usepackage{soul}

\begin{document}

\begin{frontmatter}
\title{Model-agnostic post-hoc explainability for recommender systems}

\author[inst1]{Irina Ar\'evalo\corref{cor1}}

\author[inst2]{Jose L. Salmeron}

\affiliation[inst1]{organization={School of Civil Engineering, Universidad Politécnica de Madrid},
            country={Spain}}

\affiliation[inst2]{organization={School of Engineering, CUNEF University},
            city={Madrid},
            country={Spain}}


\cortext[cor1]{Corresponding author}



\begin{abstract}
Recommender systems often benefit from complex feature embeddings and deep learning algorithms, which deliver sophisticated recommendations that enhance user experience, engagement, and revenue. However, these methods frequently reduce the interpretability and transparency of the system. In this research, we develop a systematic application, adaptation, and evaluation of deletion diagnostics in the recommender setting. The method compares the performance of a model to that of a similar model trained without a specific user or item, allowing us to quantify how that observation influences the recommender, either positively or negatively. To demonstrate its model-agnostic nature, the proposal is applied to both Neural Collaborative Filtering (NCF), a widely used deep learning-based recommender, and Singular Value Decomposition (SVD), a classical collaborative filtering technique. Experiments on the MovieLens and Amazon Reviews datasets provide insights into model behavior and highlight the generality of the approach across different recommendation paradigms.
\end{abstract}




\end{frontmatter}


\section{Introduction}
\label{sec:intro}

Recommender Systems (RS) have become a cornerstone of many online platforms and digital services, shaping user experiences and driving business performance. They are widely applied across domains such as e-commerce \cite{schafer.99}, social networks \cite{rivas.20}, digital content platforms \cite{chang.17}, and personalized learning environments. By analyzing user preferences, behaviors, and interactions, these systems deliver personalized recommendations that improve engagement and satisfaction while also increasing revenue and customer retention.

Methods for building recommender systems range from traditional collaborative filtering and content-based approaches to advanced techniques such as matrix factorization, deep learning, and hybrid models. While these achieve high accuracy, they often function as black boxes, obscuring the rationale behind their recommendations. This lack of interpretability raises concerns about transparency, fairness, and bias.

Explainability has emerged as a way to address these concerns, enhancing transparency and trust while supporting debugging and model governance. Human-interpretable explanations can increase user trust \cite{tintarev.07}, support informed decision making \cite{tintarev.072}, and encourage exploration of novel content \cite{carenini.01}. For practitioners, explanations offer diagnostic insights that guide model refinement. In practice, there is a growing demand for \textit{post hoc}, model-agnostic methods (techniques applied after training) that enable stakeholders to interrogate and validate complex RS models.

As an example, consider a small movie recommendation dataset. If a user consistently rates niche or unpopular items, their data may introduce noise into the latent space. Retraining the model without that user and observing improved Mean Average Precision (MAP) or Normalized Discounted Cumulative Gain (NDCG) would indicate a negative influence. Conversely, removing a user whose ratings are diverse and representative might sharply reduce performance, indicating a positive influence. This illustrates how deletion diagnostics can pinpoint training instances that help or hinder recommendation quality.

The main contributions of this research are two-fold:
\begin{enumerate}
\item The development of a post hoc, model-agnostic explainability method for recommender systems capable of identifying the most and least influential data points.
\item The application of this method to Neural Co\-lla\-borative Filtering and Singular Value Decomposition models, evaluated on the MovieLens and Amazon Reviews datasets, two widely used benchmarks in recommender system research.
\end{enumerate}

The remainder of the paper is organized as follows: Section~\ref{sec:relatedwork} reviews related works on explainable RS, NCF, and SVD. Section~\ref{sec:proposal} presents the deletion diagnostics method. Section~\ref{sec:exps} reports experimental results, and Section~\ref{sec:conclusions} concludes.

\section{Related works}
\label{sec:relatedwork}

\subsection{Explainable recommendations}

Explainability is a critical concern in recommender systems (RS), as these algorithms curate large volumes of content, shaping user experiences and influencing decision-making \cite{zhou.2023}. However, recommendations often embed systematic biases originating from user behavior or item characteristics \cite{deldjoo.2024, chen.2023}, which can distort outputs and amplify existing inequities \cite{mansoury.2020}. Addressing these issues is vital for transparency, fairness, and user trust.

Bias mitigation strategies typically target three stages \cite{zhou.2023}: (a) pre-processing, where data balancing or annotation reduces bias at the source; (b) in-process techniques applied during model training; and (c) \textit{post-processing} re-ranking methods \cite{li.2023, deldjoo.2024}. Beyond bias, explainability research can be divided into model-intrinsic approaches, which design inherently transparent recommenders, and model-agnostic approaches, which generate explanations after the model is trained \cite{zhang.2020}.

Inherently interpretable methods, such as content-based filtering \cite{vig.09} and neighborhood-based collaborative filtering \cite{herlocker.00, sarwar.01}, offer clear explanations that preserve fidelity; however, their performance often deteriorates in complex scenarios. Contemporary latent factor and neural models, including Neural Collaborative Filtering, demonstrate superior accuracy at the expense of transparency. Hybrid methods have been proposed to improve interpretability by aligning latent factors with external features \cite{bauman.17, lu.18, zhang.14} or by preserving neighborhood structure \cite{abdollahi.17}.

Model-agnostic techniques such as SHAP \cite{Ribeiro2016} and LIME \cite{Tohidi2024} approximate model behavior through local surrogate models or feature attribution methods, providing flexibility across different architectures but sometimes sacrificing fidelity to the original model. Influence functions \cite{Koh2017} estimate the impact of individual training points via gradient-based approximations, enabling scalable sensitivity analyses while potentially failing to capture the full effect of complex retraining dynamics. The present proposal departs from these approaches by employing deletion-based diagnostics in which the model is fully retrained without specific users or items, and the resulting changes are measured using recommendation-specific metrics such as MAP and NDCG. This methodology captures global, data-driven influence patterns that can inform model debugging, guide data curation, and support robustness evaluation.

While inherently interpretable models remain valuable for user-facing transparency, our performance-aligned, deletion-based method provides high-fidelity insights for system developers and evaluators. It complements existing approaches by revealing how individual data points shape global system behavior, bridging a gap between accuracy-driven models and interpretable, accountable recommendation pipelines.

\subsection{Neural Collaborative Filtering}

Neural Collaborative Filtering (NCF) is a neural network–based methodology within the domain of collaborative filtering (CF), a family of techniques that predict user preferences for items by leveraging historical interaction data. A widely adopted CF approach is matrix factorization, which encodes each user and each item as a real-valued latent feature vector. These vectors are dense, continuous-valued representations that capture abstract characteristics inferred during model training. The interaction between users and items is then estimated through an inner product. Although effective in practice, this linear formulation is inherently limited in its capacity to model complex, non-linear relationships between users and items.

He et al.~\cite{He.2017} introduced NCF as a framework that replaces the fixed inner product with a neural architecture capable of learning arbitrary interaction functions from data. NCF employs embedding layers for users and items and then processes them through two components: (a) Generalized Matrix Factorization (GMF), which multiplies user and item embeddings to capture linear effects, and (b) a Multi-Layer Perceptron (MLP), which concatenates the embeddings and learns non-linear patterns through multiple hidden layers. The outputs of GMF and MLP are combined in a NeuMF layer, followed by a sigmoid activation to produce the final prediction.

Compared to traditional matrix factorization, NCF can integrate diverse side information, such as item metadata or user demographics, and better model complex preference structures. It has been successfully deployed in domains including healthcare~\cite{Ponnusamy.23}, education~\cite{Mulyana_Rumaisa_2024}, tourism~\cite{marzuki.2024, wei.2025}, and real estate~\cite{venkatesh.2024}. However, this flexibility comes at a cost: as the number of users and items grows, NCF’s computational demands for training and inference increase, and the model remains a black box, offering limited interpretability.

NCF also inherits the cold-start challenge common to CF models, where new users or items with no historical data are difficult to recommend for without auxiliary features. Strategies to mitigate this include hybrid modeling with content-based methods, knowledge graph integration, transfer learning, or active exploration.

In this research, NCF is one of the two paradigms evaluated using our proposed deletion diagnostics framework, alongside the classical Singular Value Decomposition model. This dual evaluation not only demonstrates the method’s model-agnostic nature but also enables a direct comparison between a deep, non-linear architecture and a traditional matrix factorization approach in terms of influential user/item detection and the resulting impact on recommendation performance. The opacity of NCF in particular underscores the relevance of applying post-hoc, model-agnostic diagnostics to provide interpretable insights into its behavior.

\subsection{Singular Value Decomposition in RS}

Singular Value Decomposition (SVD) is a classical matrix factorization technique that has been extensively used in collaborative filtering for recommender systems \cite{koren2009matrix}. From a more formal point of view, let $A \in \mathbb{R}^{m\times n}$ be a real-valued matrix.  The SVD of $A$ is given by
\begin{equation}
A = U \Sigma V^{\top}
\label{eq:svd}
\end{equation}
where \(U \in \mathbb{R}^{m \times m}\) is an orthogonal matrix whose columns \(u_1, u_2, \dots, u_m\) are called the left singular vectors of $A$. $V \in \mathbb{R}^{n \times n}$ is an orthogonal matrix whose columns $v_1, v_2, \dots, v_n$ are called the right singular vectors of $A$. $\Sigma \in \mathbb{R}^{m \times n}$ is a diagonal matrix containing the singular values $\sigma_1, \sigma_2, \dots, \sigma_r$ (with $r = \mathrm{rank}(A)$) along the main diagonal, satisfying
\begin{equation}
    \sigma_1 \geq \sigma_2 \geq \dots \geq \sigma_r > 0.
\end{equation}

The matrices $U$ and $V$ satisfy the orthogonality conditions
\begin{equation}
U^{\top} U = I_m, \quad V^{\top} V = I_n.
\end{equation}

The singular values are related to the eigenvalues of $A^{\top} A$ and $A A^{\top}$ by $A^{\top} A = V \Sigma^{\top} \Sigma V^{\top}$ and $A A^{\top} = U \Sigma \Sigma^{\top} U^{\top}$, where $\Sigma^{\top} \Sigma$ and $\Sigma \Sigma^{\top}$ are dia\-go\-nal matrices with entries $\sigma_1^2, \dots, \sigma_r^2$. When only the non-zero singular values and corresponding singular vectors are retained, the compact or economy SVD is written as follows
\begin{equation}
A = U_r \Sigma_r V_r^{\top},
\end{equation}
where $U_r \in \mathbb{R}^{m \times r}$, $\Sigma_r \in \mathbb{R}^{r \times r}$, and $V_r \in \mathbb{R}^{n \times r}$, with $r = \mathrm{rank}(A)$.

The Moore–Penrose pseudoinverse of \(A\) can be expressed in terms of the SVD as
\begin{equation}
A^{+} = V \Sigma^{+} U^{\top},
\end{equation}
where $\Sigma^{+} \in \mathbb{R}^{n \times m}$ is obtained by taking the reciprocal of each non-zero singular value in $\Sigma$ and transposing the matrix. The SVD provides the basis for the best rank-$k$ approximation of $A$ in both the spectral norm and the Frobenius norm, as stated by the Eckart–Young–Mirsky theorem:
\begin{equation}
A_k = \arg \min_{\mathrm{rank}(B) \leq k} \|A - B\|_F,
\end{equation}
with solution
\begin{equation}
A_k = \sum_{i=1}^k \sigma_i u_i v_i^{\top}.
\end{equation}
The corresponding approximation error in the Frobenius norm is as follows
\begin{equation}
\|A - A_k\|_F = \left( \sum_{i=k+1}^r \sigma_i^2 \right)^{1/2}.    
\end{equation}

SVD factorizes the user-item interaction matrix into the product of three lower-dimensional matrices, capturing latent factors that represent user preferences and item characteristics. By projecting both users and items into a shared latent space, SVD estimates missing ratings through the dot product of corresponding user and item vectors. This method gained prominence through its success in the Netflix Prize competition, where it became a core component of many high-performing hybrid recommendation approaches.

Singular Value Decomposition presents several advantages, including computational efficiency for datasets of moderate size, the production of interpretable latent factors, and strong performance in scenarios characterized by dense ratings. Nevertheless, the linear nature of SVD restricts its capacity to model complex, non-linear relationships between users and items, particularly when compared to deep learning approaches such as Neural Collaborative Filtering. Despite this limitation, SVD continues to serve as a robust baseline within recommender system research and is widely adopted owing to its simplicity, scalability, and reliable performance across a variety of practical applications.

This research uses Singular Value Decomposition alongside Neural Collaborative Filtering to prove the model-agnostic properties of the proposed deletion diagnostics framework. By performing the same influence analysis on SVD, we evaluate the method’s ability to consistently identify influential users and items across fundamentally different recommendation paradigms. This comparison reinforces the approach’s versatility and applicability in diverse contexts.

\subsection{Popularity bias}

Popularity bias is a well-documented challenge in RS, characterized by algorithms disproportionately favoring popular items, often at the expense of recommending diverse or niche content. This bias can reduce the variety of recommendations presented to users, negatively impacting user satisfaction and potentially reinforcing existing disparities among items or providers. In deep learning-based models such as Neural Collaborative Filtering, popularity bias may arise implicitly, as the model optimizes for overall accuracy, which frequently correlates with recommending widely liked or frequently interacted-with items \cite{abdollahpouri2017controlling}.

The application of deletion diagnostics offers a novel perspective to investigate popularity bias. By quantifying the influence of individual users or items on recommendation outcomes, it becomes possible to identify whether certain popular items exert a disproportionate effect on the model's behavior. For example, the removal of a popular item causing significant changes in recommendations or model performance may reveal an excessive dependence on such items, indicating the presence of popularity bias.

Furthermore, the detailed influence scores gene\-ra\-ted through deletion diagnostics can inform strate\-gies aimed at mitigating popularity bias \cite{steck2018calibrated}. Identifying which popular items dominate the model's decision-making process enables practitioners to imple\-ment reweighting schemes or regularization techniques to encourage a more balanced recommendation distribution favoring diverse content. These insights contribute to enhancing fairness and improving user experience by promoting a wider variety of items aligned with individual user preferences.

\section{Methodological proposal}
\label{sec:proposal}
\subsection{Proposed method}

This research introduces deletion diagnostics as a post-hoc explainability technique that quantifies the influence of each user or item on the performance of a recommender system. The method identifies the most influential training instances, defined as those whose removal induces substantial changes in evaluation metrics, either positive or negative. Analyzing these instances yields deeper insight into the learning dynamics of recommender systems, elucidating which user or item characteristics drive predictions and highlighting potential areas for debugging or refinement.

In contrast to many explainability approaches in recommender systems, such as SHAP, LIME, attention mechanisms, or counterfactual explanations, which typically focus on local, instance-level explanations or rely on surrogate models, the proposed method directly measures the global impact of removing individual data points. This deletion-based analysis naturally aligns with recommendation-specific metrics such as Mean Average Precision and Normalized Discounted Cumulative Gain, providing a performance-centered perspective on influence. Rather than approximating effects via gradients or simplified models, the approach captures the full retraining impact, thereby offering high-fidelity insights into data quality, robustness, and fairness.

To validate the method, we apply it to two contrasting recommendation algorithms: Neural Collaborative Filtering, a deep learning model that captures complex non-linear user–item interactions but offers little transparency; and Singular Value Decomposition, a classic latent factor model with simpler, more interpretable representations. Evaluating both demonstrates the method's applicability across architectures and reveals differences in how neural and matrix factorization models leverage user and item data.

By bridging the interpretability gap between accuracy-oriented models and actionable developer insights, this approach offers a flexible framework for model auditing, robustness assessment, and fairness analysis. Its direct, perturbation-based nature ensures that findings are grounded in measurable performance changes, making it a practical tool for improving recommender systems in diverse settings.

\subsection{Deletion diagnostics for influence estimation}
\label{subsec:deletion_diagnostics}

The influence of a user or item $i$ is computed as:

\begin{equation}
    \text{Influence}^{(-i)} = \text{eval} - \text{eval}^{(-i)}
\end{equation}

\noindent where $\text{eval}$ is the evaluation metric on the original model and $\text{eval}^{(-i)}$ is the same metric after retraining without $i$.

\begin{algorithm}
\small
\DontPrintSemicolon
\SetAlgoLined
\KwData{Train data ($\mathcal{X}$) and MAP of the NCF}
\KwResult{Dictionary of differences}

\For{\text{User} $u=1$ \KwTo $n = \#\text{users}$}{
    \textbf{Data generation:} Generate the data without participant $u$ and split train/test\;
    $\mathcal{X}^{(-u)} = \mathcal{X}\backslash\{u\}$\;
    \textbf{Train model:} Train NCF model with train data from $\mathcal{X}^{(-u)}$ to obtain NCF$^{(-u)}$\;
    \textbf{Predictions and metrics} for MAP$^{(-u)}$\;
    \textbf{Append to differences:} MAP $-$ MAP$^{(-u)}$
}
\caption{Influential observations for users}\label{alg:inf_user}
\end{algorithm}

\begin{algorithm}
\small
\DontPrintSemicolon
\SetAlgoLined
\KwData{Train data ($\mathcal{X}$) and MAP of the NCF}
\KwResult{Dictionary of differences}

\For{\text{Item} $i=1$ \KwTo $m = \#\text{items}$}{
    \textbf{Data generation:} Generate the data without item $i$ and split train/test\;
    $\mathcal{X}^{(-i)} = \mathcal{X}\backslash\{i\}$\;
    \textbf{Train model:} Train NCF model with train data from $\mathcal{X}^{(-i)}$ to obtain NCF$^{(-i)}$\;
    \textbf{Predictions and metrics} for MAP$^{(-i)}$\;
    \textbf{Append to differences:} MAP $-$ MAP$^{(-i)}$
}
\caption{Influential observations for items}\label{alg:inf_item}
\end{algorithm}

Algorithms \ref{alg:inf_user} and \ref{alg:inf_item} detail the computation for users and items, respectively.

\subsection{Evaluation metrics}
\label{subsec:metrics}

This research focuses on ranking- and accuracy-based metrics commonly used to evaluate RS quality \cite{jadon2024, zangerle.2022}:

\begin{itemize}
    \item Mean Average Precision (MAP): mean of the average precision at each point a relevant item is retrieved, averaged over all users. 
    \item Mean Average Precision@k (MAP@k): evaluates the average precision at all relevant ranks within the list of top $k$ recommendations. 
    \item Normalized Discounted Cumulative Gain (n-DCG): evaluates how well the predicted ranking of items corresponds to the ideal ranking considering the relevance of each item.
    \item Precision@k: proportion of recommended items in the top-$k$ set that are relevant to the user. 
    \item Recall@k: proportion of relevant items found in the top-$k$ recommendations. 
    \item Explained Variance: proportion of variance in the target explained by the independent variables in the model
    \item Mean Absolute Error (MAE): average magnitude of errors in a set of predictions, without considering their direction.
\end{itemize}

While the method is metric-agnostic, MAP is selected for this research as it emphasizes higher-ranked relevant items. Formally:

\begin{equation}
    \text{MAP} = \frac{1}{|\mathcal{U}|}\sum_{u = 1}^{|\mathcal{U}|}\left(\frac{1}{|\mathcal{R}_u|}\sum_{k = 1}^{|\mathcal{S}_u|} \text{P@}k\cdot\text{I}_u(k)\right)
\label{eq:map}
\end{equation}

\subsection{Computational Cost and Scalability}
\label{subsec:cost}

This approach is computationally demanding, as it requires retraining the model for each user or item. While gradient-based or surrogate-model methods (e.g., SHAP, LIME) are faster, they rely on approximations. Deletion diagnostics, in contrast, capture actual performance changes, ensuring fidelity at the expense of higher cost.

Table~\ref{tab:time_complexity_comparison} compares the computational cost of the proposed method with other popular explainability techniques. Perturbation-based methods such as SHAP and LIME avoid full retraining but may still require many predictions, which can be costly for large models. Thus, although deletion diagnostics are more resource-intensive, their flexibility and applicability to black-box recommenders make them attractive where model-specific assumptions cannot be met.

\begin{table}[ht]
\centering
\footnotesize
\caption{Time complexity for SOTA explai\-na\-bi\-lity methods}
\begin{tabular}{p{1.8cm}|l|p{3.2cm}}
\hline
\textbf{Method} & \textbf{Complexity} & \textbf{Notes} \\ \hline\hline
Deletion Diagnostics (proposed) & $O(n \times T_{\mathrm{train}})$ & 
Requires retraining once per instance (user/item); fully model-agnostic. \\\hline
SHAP (sampling-based) & $O(m \times P)$ & 
$m$ = number of perturbed samples, $P$ = prediction cost, no full retraining. \\\hline
LIME & $O(m \times P)$ & 
Similar to SHAP but fits a local surrogate model; no retraining required. \\ \hline
\end{tabular}
\label{tab:time_complexity_comparison}
\end{table}

Scalability can be improved in practice using parallelization, subsampling, or focusing on top-$K$ candidates for influence assessment. Parallel retraining can distribute workload across multiple computing units, sampling can target representative subsets of users or items, and lightweight proxy models or partial retraining methods can approximate influence scores with reduced cost. These strategies are promising avenues for enabling real-time or production-scale deployment.

Beyond computational considerations, this method primarily serves model developers, data scientists, and system maintainers by enabling:

\begin{itemize}
    \item Model debugging: Identifying problematic or overly influential users/items that may skew recommendations or introduce bias.
    \item Data curation: Guiding the removal or re-weighting of noisy, adversarial, or outlier data points to improve overall system robustness and fairness.
    \item Performance auditing: Evaluating which subsets of data support or hinder model generalization and tuning the training process accordingly.
    \item Algorithmic transparency: Offering insights into model dynamics without requiring access to internal parameters or architecture, which is especially valuable when dealing with black-box or proprietary recommenders.
\end{itemize}

While these benefits are developer-focused, end-users may indirectly gain from improved fairness, reliability, and trust in recommendations. Communicating that such influence analyses are conducted can further reinforce user confidence. The framework remains model-agnostic and is applicable to other recommender families, including matrix factorization, autoencoders, and hybrid systems, making its extension to diverse models and datasets a promising direction for future research.

\section{Experimental approach}
\label{sec:exps}

\begin{figure}[ht!]
    \centering
    \includegraphics[width=0.80\linewidth]{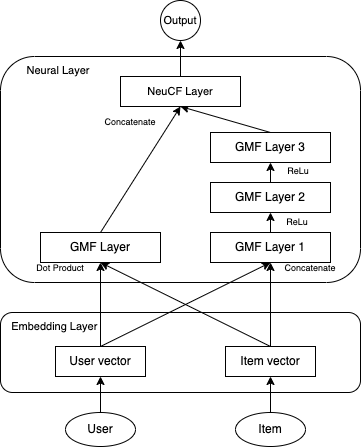}
    \caption{Architecture of the NCF}
   \label{fig:arch}
\end{figure}

This proposal has been tested using two datasets: MovieLens and Amazon Reviews. The experimental approach begins with a 75\%/25\% train/test split, with 75\% of the data in the training set. For the MovieLens dataset, two recommendation models are constructed: a Neural Collaborative Filtering model and a Singular Value Decomposition model, allowing for a comparative evaluation between a deep learning-based approach and a classical matrix factorization method. For the Amazon Reviews dataset, only the NCF model is tested due to the binary nature of the data. The NCF model generates a latent space with a dimension of 4, and its MLP consists of three hidden layers with 16, 8, and 4 neurons respectively. It uses a learning rate of $1\mathrm{e}{-3}$, a batch size of 256, and is trained for 10 epochs. These hyperparameters are typical in the literature, but it should be noted that this proposal is agnostic to both the hyperparameters and the model architecture, allowing it to be applied to other approaches such as SVD. A sketch of the architecture of the NCF model can be found in Figure \ref{fig:arch}.

The proposed deletion diagnostics method goes beyond simply ranking the influence of users and items. By linking influence scores with concrete operational decisions, the method enables systematic interventions in data curation, model training, platform design, and product strategy. These subsections expand on MovieLens and Amazon case studies with domain-specific, actionable scenarios.

\subsection{MovieLens - NCF}

The first experiment is conducted on the 100K MovieLens dataset \cite{Harper.2015}, a publicly available dataset that contains movie ratings that is commonly used to evaluate collaborative filtering algorithms, such as those in \cite{rendle.2020, Kammoun.2022, fan.2024, chen.2024}.  This dataset contains 100K ratings in a set of tuples with information about the user, the movie, the timestamp of the rating, and the rating provided by the user for the movie. The rating is a number from 1 to 5. A summary of the details of the dataset can be found in Table \ref{tab:eda}.

\begin{table}[ht]
    \centering
    \caption{100K MovieLens Dataset metadata}
    \begin{tabular}{l|r}
    \hline
        Metric      & 100K MovieLens \\\hline\hline
        \# Ratings   & 100,000   \\
        \# Items     & 1,682       \\
        \# Users     & 943  \\
        Density     & 0.063 \\
        Average \# ratings/item & 59.45 \\
        Average \# ratings/user & 106.05 \\
        Min \# ratings/item & 1 \\
        Min \# ratings/user & 20 \\
        Max \# ratings/item & 583 \\        
        Max \# ratings/user & 737 \\
        
        Average rating & 3.53\\ \hline
    \end{tabular}
    \label{tab:eda}
\end{table}

The evaluation metrics for the NCF model trained with the 100K MovieLens dataset are shown in Table \ref{tab:metrics}.

\begin{table}[ht]
    \centering
    \caption{Performance metrics for the NCF model on the MovieLens dataset}
    \begin{tabular}{l|c}
    \hline
    Metric & Value \\\hline\hline
    MAP & 0.047070 \\
    MAP@K & 0.100803 \\
    NDCG & 0.193374 \\
    Precision@K & 0.173595 \\
    Recall@K & 0.096695\\ 
    Explained Var & 0.055031\\
    MAE & 2.984612\\
    \hline
    \end{tabular}
    \label{tab:metrics}
\end{table}

\begin{figure}
    \centering
    \includegraphics[width=0.85\linewidth]{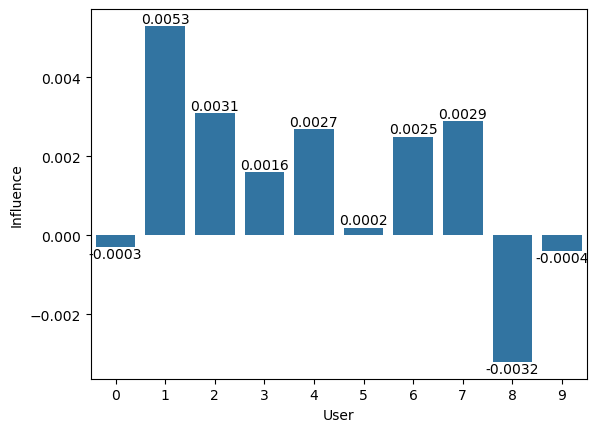}
    \caption{Result of influences for 10 first users}
    \label{fig:plot2}
\end{figure}

To find the most and least influential users for this model one should follow the steps in Algorithm \ref{alg:inf_user}. The results for the first 10 users can be found in Figure \ref{fig:plot2}. With this approach the authors find that user 2 is the user that most positively influences this model, since the difference between the MAP of the global model and the MAP of the model without user 2 reaches its maximum. 

\begin{figure}
    \centering
    \includegraphics[width=0.90\linewidth]{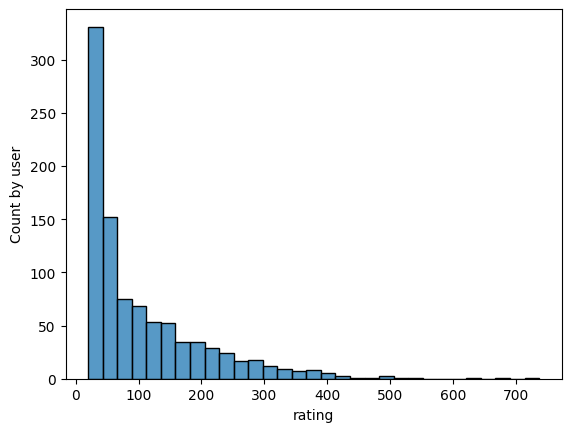}
    \caption{Histogram of number of ratings by user}
    \label{fig:plot1}
\end{figure}

This user has provided 62 ratings, with an average of 3.70 and a standard deviation of 1.03. Although it might seem that the user with the most ratings would be the most influential, this is not always true, as Figure \ref{fig:plot1} shows. In terms of the average and standard deviation of their ratings, they are quite typical. However, this information could be useful for designing a clustering strategy based on their characteristics or those of the items they have rated. 

The user who has the most negative impact on the model is user 273. This is because the difference in the MAP between the original model and the model trained without user 273 is minimal. In fact, removing this user slightly improves the MAP metric for the overall model. Table \ref{tab:neg_infl} shows a summary of the ratings made by the five users that have a greater negative impact on the model. 

In this case, 4 out of 5 users have very few ratings, close to the minimum number, but one does not. In terms of average ratings, there is some variation. Specifically, two participants have average ratings similar to those of the most positively influential user (User 2), though with a smaller standard deviation.

\begin{table}[ht]
    \centering
        \caption{Summary of ratings from most negatively influential users}
    \begin{tabular}{c|c|c|c}
    \hline
        User ID & \# ratings & Average rating & Std. rating \\\hline\hline
        273 & 22  & 3.64 & 0.73\\
         17 & 28  & 3.04 & 1.14\\
        775 & 28  & 3.79 & 0.79\\
        311 & 294 & 3.80 & 0.87\\
        583 & 27  & 4.44 & 0.70\\ \hline
    \end{tabular}
    \label{tab:neg_infl}
\end{table}

As a way of testing the effectiveness of this proposal and understand how the most and least influential users affect the model, the authors have trained the same architecture without the most and least influential users according to this proposal. The performance metrics of the model without the 10 most influential users can be found in Table \ref{tab:metrics_most_infl_users}. In this table we can find the name of the metric, the value without the most influential participant (Value wo. most influential), the original value of the metric, and the difference.

\begin{table}[ht]
    \centering
        \caption{Performance metrics for the NCF model without the 10 most influential users}
\begin{tabular}{l|c|c|c}
    \hline
     & Value wo. & Original & \\
     Metric &  most  & Value & Difference\\
     & influential &  &\\\hline\hline
    MAP &       0.045797 & 0.047070 &  \textbf{-2.70\%} \\
    MAP@K &     0.100303 & 0.100803 &  \textbf{-0.50\%} \\
    NDCG  &     0.200036 & 0.193374 & 3.45\% \\
  Precision@K & 0.164962 & 0.173595 &  \textbf{-4.97\%} \\
     Recall@K & 0.092232 & 0.096695 &  \textbf{-4.62\%} \\
Explained Var & 0.056246 & 0.055031 & 2.21\% \\
          MAE & 3.013710 & 2.984612 & \textbf{0.97\%} \\
    \hline
    \end{tabular}
    \label{tab:metrics_most_infl_users}
\end{table}

It would be expected that removing the most influential users from the model would lead to worse performance compared to the original model, reflected in lower values for MAP, MAP@k, NDCG, Precision@k, Recall@k, and Explained Variance, as well as a higher MAE. In this case, Table \ref{tab:metrics_most_infl_users} shows that 5 out of 7 metrics are either worse than or equal to those in Table \ref{tab:metrics}, including a nearly 5\% decrease in Precision@k and Recall@k, and a nearly 1\% increase in MAE. 

The performance metrics of the NCF without the 10 least influential users are shown in Table \ref{tab:metrics_least_infl_users}, that follows a similar format as Table \ref{tab:metrics_most_infl_users} with the value of the metric without the most influential user (Value wo. most influential). 

\begin{table}[ht]
    \centering
    \caption{Performance metrics for the NCF model without the 10 least influential users}
    \begin{tabular}{l|c|c|c}
    \hline
     & Value wo. & Original & \\
     Metric &  least  & Value & Difference\\
     & influential &  &\\\hline\hline
    MAP &       0.048451 & 0.047070 & \textbf{2.93\%} \\
    MAP@K &     0.119442 & 0.100803 & \textbf{18.49\%} \\
    NDCG  &     0.219651 & 0.193374 & \textbf{13.59\%} \\
  Precision@K & 0.202680 & 0.173595 & \textbf{16.75\%} \\
     Recall@K & 0.099090 & 0.096695 & \textbf{2.48\%} \\
Explained Var & 0.063860 & 0.055031 & \textbf{16.04\%} \\
          MAE & 2.973177 & 2.984612 & \textbf{-0.38\%} \\
    \hline
    \end{tabular}
    \label{tab:metrics_least_infl_users}
\end{table}

In this case, the expectation was that removing these 10 users would lead to an improvement in the metrics. However, the table shows that all metrics actually improve, including nearly a 20\% increase in MAP@k and Precision@k. It is important to note that the metrics are more strongly affected by the removal of the least influential users than by the removal of the most influential ones. Nonetheless, the difference in MAP is similar in both models (without the most and least influential participants), likely because MAP was used as optimization metric for deletion diagnostics.

The results in Tables \ref{tab:metrics_most_infl_users} and \ref{tab:metrics_least_infl_users} demonstrate that the deletion diagnosis successfully identified both the most and, especially, the least influential users for this model. The improvement in performance following the removal of the least influential users suggests that these users contribution is minimal or even detrimental signal during training. Their sparse, redundant, or non-generalizable interactions may introduce noise into the learning process, leading to poorer generalization. Removing them results in a smoother latent space, reduced model variance, and optimization more tightly aligned with the evaluation objective. These insights highlight the value of influence-based diagnostics not just for explanation, but also for data curation and model refinement.

\begin{figure}
    \centering
    \includegraphics[width=0.85\linewidth]{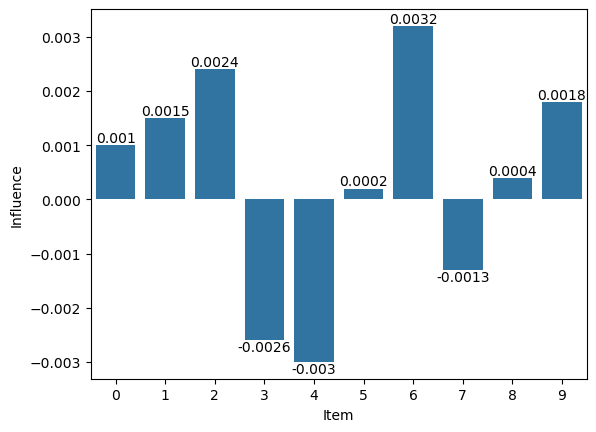}
    \caption{Result of influences for 10 first items}
    \label{fig:plot3}
\end{figure}

\begin{figure}
    \centering
    \includegraphics[width=0.90\linewidth]{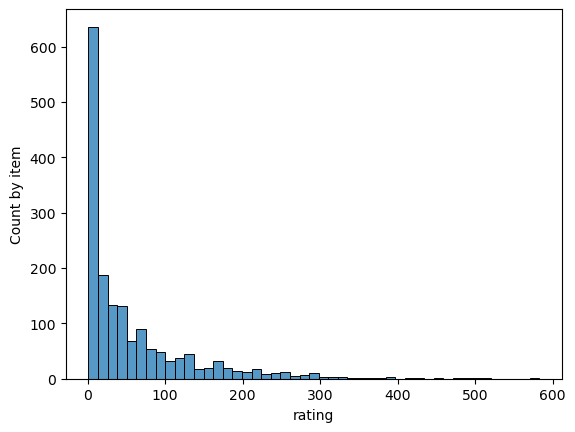}
    \caption{Histogram of number of ratings by item}
    \label{fig:plot4}
\end{figure}

Regarding to items, Algorithm \ref{alg:inf_item} outlines the process to find the most influential items for the recommender. The diffe\-ren\-ces in MAP for the first 10 items can be found in Figure \ref{fig:plot3}. The item that contributes most positively to the recommendation is item 1227, a movie with 8 reviews with an average ra\-ting of 2.4 and a standard deviation of 1.06. Again, one could expect the most influential item to have a large amount of reviews, but movie 1227 is close to the 0.25 quantile. Figure \ref{fig:plot4} displays the histogram of ratings by item.

The five items that have the most effect in the MAP of the recommender in a negative way are displayed in Table \ref{tab:neg_infl_item}. For the case of items, again one could expect these items to be the ones with least number of reviews, and indeed, as in the case of users, there are several items with very few ratings and one movie with much more occurrences in the dataset. The mean rating and its deviation are also more diverse, but affected by the low number of ratings (as shown by movie 1342, with only two ratings and a standard deviation of 2.12). 

\begin{table}[ht]
    \centering
    \caption{Summary of ratings from most negatively influential items}
    \begin{tabular}{c|c|c|c}
    \hline
        Item ID & \# ratings & Average rating & Std. rating \\\hline\hline
         905 & 27  & 3.22 & 0.93 \\
        1342 &  2  & 2.50 & 2.12\\
        1518 & 12  & 3.00 & 0.85\\
        1380 &  6  & 2.33 & 1.03\\
         482 & 128 & 3.98 & 0.97\\\hline
    \end{tabular}
    \label{tab:neg_infl_item}
\end{table}

To better understand how the most and least influential items and users impact the model, the same model can be trained without the 10 most influential items and without the 10 least influential items. The performance results for the model without the 10 most influential items can be found in Table \ref{tab:metrics_most_infl_items}, with the same structure as Table \ref{tab:metrics_most_infl_users}.

\begin{table}[ht]
    \centering
    \caption{Performance metrics for the NCF model without the 10 most influential items}
    \begin{tabular}{l|c|c|c}
    \hline
     & Value wo. & Original & \\
     Metric &  most  & Value & Difference\\
     & influential &  &\\\hline\hline
          MAP   & 0.046185 & 0.047070 & \textbf{-1.88\%}\\
          MAP@K & 0.102364 & 0.100803 & +1.55\%\\
          NDCG  & 0.192648 & 0.193374 & \textbf{-0.38\%}\\
    Precision@K & 0.173171 & 0.173595 & \textbf{-0.24\%}\\
       Recall@K & 0.094898 & 0.096695 & \textbf{-1.86\%}\\ 
  Explained Var & 0.058646 & 0.055031 & +6.57\%\\
 MAE & 2.986054 & 2.984612 & \textbf{+0.05\%}\\
    \hline
    \end{tabular}
    \label{tab:metrics_most_infl_items}
\end{table}

As in the previous case, it is expected that removing the most influential items would lead to a decline in the model's performance. In this case 5 out of 7 metrics show a worse performance after removing the 10 most influential items, although the percentage reduction is not as sharp as in the case of users. This suggests that users, or at least the 10 most influential users, play a more significant role in the model's performance than the items.

The performance results for the model without the 10 least influential items can be found in Table \ref{tab:metrics_least_infl_items}, again with the same structure as Table \ref{tab:metrics_least_infl_users}.

\begin{table}[ht]
    \centering
    \caption{Performance metrics for the NCF model without the 10 least influential items}
    \begin{tabular}{l|c|c|c}
    \hline
     & Value wo. & Original & \\
     Metric &  least  & Value & Difference\\
     & influential &  &\\\hline\hline
              MAP & 0.048589 & 0.047070 & \textbf{+3.23\%}\\
            MAP@K & 0.102536 & 0.100803 & \textbf{+1.72\%}\\
            NDCG  & 0.193554 & 0.193374 & \textbf{+0.09\%}\\
      Precision@K & 0.172534 & 0.173595 & -0.61\%\\
         Recall@K & 0.098822 & 0.096695 & \textbf{+2.20\%}\\ 
    Explained Var & 0.057276 & 0.055031 & \textbf{+4.08\%}\\
              MAE & 2.976582 & 2.984612 & \textbf{-0.27\%}\\
    \hline
    \end{tabular}
    \label{tab:metrics_least_infl_items}
\end{table}

Once again, removing the 10 least influential items improves all performance metrics except one, with a +4.08\% increase in explained variance and a +3.23\% increase in MAP. Similar to the case with users, removing the least influential items has a more significant impact on the model's performance than removing the most influential items.

In the MovieLens experiment, removing the 10 least influential users resulted in large performance gains, with MAP@K increasing by +18.49\% and Precision@K by +16.75\%. This finding indicates that certain user profiles—often characterized by sparse, inconsistent, or noisy rating patterns—are detrimental to model generalization. Practical actions informed by these findings include:
\begin{itemize}
    \item Data filtering: Down-weighting or removing low-impact users before retraining to reduce variance in the learned latent space. This is particularly valuable when curating datasets for periodic retraining cycles.
    \item Adaptive weighting: Implementing influence-weighted loss functions where high-impact users have stronger gradients during optimization.
    \item Targeted enrichment: Designing engagement campaigns to encourage low-impact users to interact with more diverse items, improving the breadth of training data.
    \item Segment-specific tuning: Creating separate model segments where high- and low-impact user profiles are treated differently, improving personalization in both clusters.
\end{itemize}

The item-level analysis revealed that item 1227, with only 8 reviews, was disproportionately beneficial to MAP, while some heavily reviewed items had negligible or negative effects. This opens avenues for:
\begin{itemize}
    \item Catalog curation: Retaining or highlighting items with strong positive influence, even if their popularity is modest, as they contribute to the diversity and accuracy of recommendations.
    \item Inventory pruning: Identifying items with consistent negative influence for removal from recommendation pools or reduced visibility in high-traffic feeds.
    \item Cold-start optimization: Studying characteristics of high-impact, low-interaction items to inform onboarding and promotion strategies for new catalog entries.
    \item Exposure control: Balancing visibility of low-impact items to minimize training inefficiencies caused by redundant or uninformative interactions.
\end{itemize}

\subsection{Amazon reviews - NCF}

The second dataset is the Amazon Reviews dataset, \cite{mcauley.2013} another publicly accessible dataset with product, in particular Electronics, reviews. This is a binary dataset: instead of ratings, this dataset contains information about products that users have reviewed. Therefore, it is assumed that each row represents a rating of 1 and we need negative samples (that is, pairs user/product that have not been reviewed) that will have a rating of 0. A summary of the characteristics of the dataset can be found in Table \ref{tab:eda_amazon}.

\begin{table}[ht]
    \centering
    \caption{Amazon Reviews Dataset metadata}
    \begin{tabular}{l|r}
    \hline
        Metric      & Amazon Reviews \\\hline\hline
        \# Ratings   & 168,919   \\
        \# Items     & 9,754       \\
        \# Users     & 15,551  \\
        Density     & 3.67e-05 \\
        Average \# ratings/item & 29.98 \\
        Average \# ratings/user & 17.16 \\
        Min \# ratings/item & 10 \\
        Min \# ratings/user & 10 \\
        Max \# ratings/item & 1376 \\        
        Max \# ratings/user & 317 \\ \hline
    \end{tabular}
    \label{tab:eda_amazon}
\end{table}

This dataset is significantly sparser than the MovieLens dataset, with users contributing far fewer reviews. The evaluation metrics for the NCF model trained with this dataset are shown in Table \ref{tab:metrics_netflix}.

\begin{table}[ht]
    \centering
    \caption{Performance metrics for the NCF model on the Amazon Reviews dataset}
    \begin{tabular}{l|c}
    \hline
    Metric & Value \\\hline\hline
    MAP & 0.000423 \\
    MAP@K & 0.000424 \\
    NDCG & 0.000961 \\
    Precision@K & 0.000557 \\
    Recall@K & 0.001113\\ 
    Explained Var & -0.033642\\
    MAE & 0.013208\\
    \hline
    \end{tabular}
    \label{tab:metrics_netflix}
\end{table}

\begin{figure}
    \centering
    \includegraphics[width=0.85\linewidth]{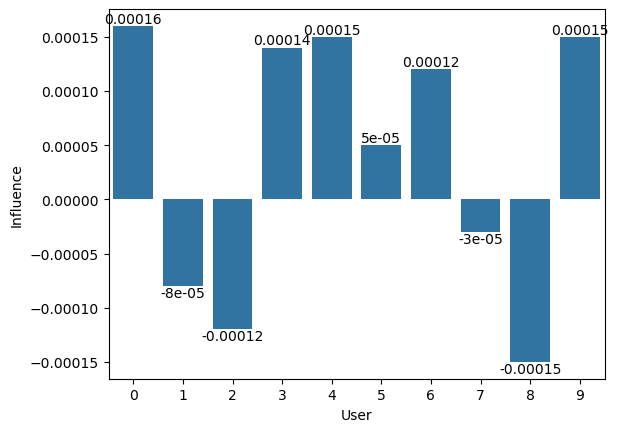}
    \caption{Result of influences for 10 first users}
    \label{fig:inf_users_amazon}
\end{figure}

The influence of the first 10 users, according to algorithm \ref{alg:inf_user}, is shown in Figure \ref{fig:inf_users_amazon}.

It is clear that the first participant is the one that most positively influences the model. This participant has 17 reviews, close to the average and far from the maximum of 317 ratings. 

Regarding the most and least influential users for this model, user 5760, with their 30 reviews, has the most predictive impact, since the difference between the MAP for the models with and without this user is maximal. The least influential user is 3248, as the difference in performance between the models with and without this user is not only minimal but also negative. Therefore, the performance of the recommender system improves when this user and their 10 reviews are removed from the dataset. Again, this means that the number of reviews is not the most important characteristic of the reviewer, but the underlying information about the items that the model is learning.

\begin{figure}
    \centering
    \includegraphics[width=0.85\linewidth]{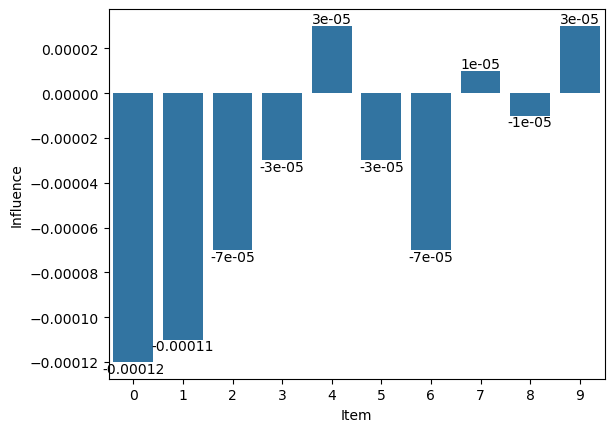}
    \caption{Result of influences for 10 first items}
    \label{fig:inf_items_amazon}
\end{figure}

\begin{figure}[t]
    \centering
    \includegraphics[width=0.85\linewidth]{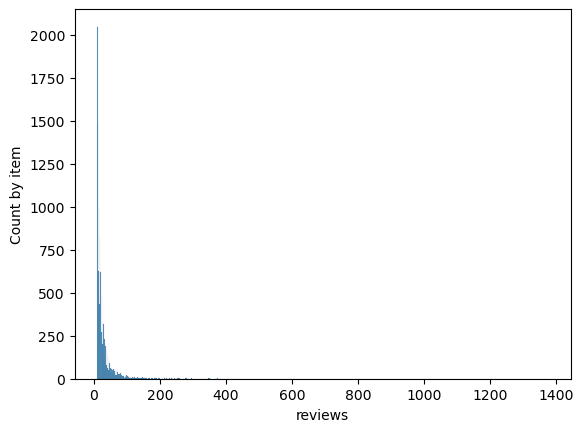}
    \caption{Histogram of reviews per item}
    \label{fig:hist_items_amazon}
\end{figure}

With respect to the influence of items, the first 10 influences, computed with algorithm \ref{alg:inf_item}, are shown in Figure \ref{fig:inf_items_amazon}. In this case, the most influential item within the first 10 is the fifth item that has 141 reviews, far above the average number of reviews per item as Figure \ref{fig:hist_items_amazon} proves. The most influential item according to the algorithm \ref{alg:inf_item} is item 9460 with 11 reviews, while the least influential is 4408 with 20 reviews. 

In this dataset, where the sparsity is extreme, the deletion diagnostics identified a single low-impact user (10 reviews) whose removal improved performance. Such insights are critical when data collection is expensive, particularly across several key domains: 

\begin{itemize}
    \item Noise reduction in sparse data: Identifying outlier profiles whose ratings deviate systematically from the broader population trend.
    \item Reviewer credibility scoring: Incorporating influence scores into trust metrics for user-generated content, particularly for platforms with verified and unverified reviewers.
    \item Dynamic weighting in live systems: Adjusting the influence of users/items in real time for online learning setups to stabilize recommendations.
\end{itemize}

Beyond improving accuracy, the method also carries strategic implications for different stakeholders. For system developers, influence diagnostics can help prioritize model retraining, identify data augmentation opportunities, and perform robustness checks to ensure stable performance. For product managers, the insights can guide catalog management decisions, shape promotion strategies, and inform user engagement initiatives. For trust and safety teams, the technique supports the detection of anomalous rating behaviors that may indicate spam, bias, or coordinated manipulation, thereby enhancing the integrity of the recommendation system.

Ultimately, deletion diagnostics are not purely descriptive but form a decision-support tool for managing both the quality of data and the integrity of recommendations. The framework systematically translates statistical influence patterns into operational strategies that enhance both the accuracy and the robustness of recommender systems in diverse domains.

\subsection{MovieLens - SVD}

To further validate the model-agnostic nature of the proposed deletion diagnostics framework, we extended the experimental evaluation to include Singular Value Decomposition, a classical and widely adopted collaborative filtering technique. Unlike NCF, which relies on non-linear transformations and deep learning architectures to model complex user–item interactions, SVD factorizes the rating matrix into lower-dimensional latent spaces for users and items, enabling efficient and interpretable recommendations. By applying the same influence analysis methodology to SVD, we assess whether the proposed approach consistently identifies influential users and items across fundamentally different recommendation paradigms, thereby reinforcing its generality and applicability beyond neural models.

The evaluation metrics for the SVD model trained with the 100K MovieLens dataset are shown in Table~\ref{tab:metrics_SVD}.

\begin{table}[ht]
    \centering
    \caption{Performance metrics for the SVD model on the MovieLens dataset}
    \begin{tabular}{l|c}
    \hline
    Metric & Value \\\hline\hline
    MAP &   0.610440 \\
    MAP@K & 1.000000 \\
    NDCG &  1.000000 \\
    Precision@K & 0.878556 \\
    Recall@K &      0.610440\\ 
    Explained Var & 0.287167\\
    MAE &     0.746868\\
    \hline
    \end{tabular}
    \label{tab:metrics_SVD}
\end{table}

The performance metrics reveal substantial differences between the NCF and SVD models on the MovieLens dataset. SVD dramatically outperforms NCF across nearly all evaluation measures, achieving perfect or near-perfect scores in MAP@K and NDCG, and very high values in Precision@K and Recall@K. It also records a much lower MAE, indicating more accurate rating predictions. While these results underscore the strong inductive bias of SVD for factorizing dense rating matrices, the attainment of perfect scores may also be influenced by specific characteristics of the dataset, particularly the relatively small size and limited diversity of MovieLens~100K. Such factors can advantage simpler, well-suited models such as SVD. In contrast, NCF, with its non-linear architecture, requires richer or more complex data to fully leverage its capacity, and may underperform on smaller datasets with limited interaction diversity. Additionally, NCF is more sensitive to hyperparameters, optimization variance, and overfitting in sparse settings, further widening the performance gap in favor of SVD for this benchmark.

For this model, the most positively influential participant (meaning that removing this user results in the largest drop in the performance metric) is user~50, who has a mean rating of 3.54 across 24 reviews. In contrast, the most negatively influential participant is user~309, with 28 reviews and a mean rating of 2.86.

The user ID, number of ratings, average rating, and standard deviation of the ratings for the five most positively influential participants are detailed in Table~\ref{tab:neg_infl_svd}.

\begin{table}[ht]
    \centering
        \caption{Summary of ratings from most positively influential users - SVD}
    \begin{tabular}{c|c|c|c}
    \hline
        User ID & \# ratings & Average rating & Std. rating \\\hline\hline
        50 &  24   &  3.54 & 1.38\\
        294 & 150  &  3.47 & 0.95\\
        1 &   272  &  3.61 & 1.26\\
        100 &  59  &  3.07 & 1.05\\
        258 & 23   &  3.74 & 1.39\\ \hline
    \end{tabular}
    \label{tab:neg_infl_svd}
\end{table}

These results further confirm that the deletion diagnostics framework is capable of identifying disproportionately influential users and items across both simple, interpretable models like SVD and more complex, non-linear architectures like NCF. This cross-paradigm consistency strengthens the claim that the method is both model-agnostic and broadly applicable within the recommender systems domain.

\section{Conclusions}
\label{sec:conclusions}

This research introduces a model-agnostic, post-hoc diagnostic framework for recommender systems, designed to quantify the influence of individual users and items on the performance of collaborative filtering models. The method operates independently of model architecture and applies to any recommender where performance can be re-evaluated after user or item removal. Its applicability is demonstrated on Neural Collaborative Filtering and Singular Value Decomposition for the MovieLens 100K dataset, and on NCF for the Amazon Reviews dataset, using MAP, MAP@K, NDCG, and related metrics to evaluate both global and perturbed models.

The experiments reveal that certain users and items act as high-leverage points in the learned latent space. Removing the most influential users degrades performance, while removing the least influential ones often yields measurable gains—for example, in MovieLens, MAP@K improved by 18.49\% after removing the 10 least influential users. These results confirm that some interactions contribute minimal or even detrimental signal, and that their removal can reduce variance and improve generalization.

While the method incurs non-trivial computational costs due to repeated retraining, it remains feasible for medium-scale datasets, and can be parallelized to improve scalability. The framework provides actionable, global-level insights that support model debugging, targeted data curation, and fairness auditing, without requiring access to gradients or internal model details—making it suitable for black-box recommenders.

Limitations include high runtime complexity on large-scale datasets, lack of direct user-facing explanations, and reliance on the chosen evaluation metric to reflect recommendation quality. Future research will explore approximation algorithms, parallelization strategies, and extensions to additional recommender architectures, enabling deployment in larger or real-time settings.

Overall, the proposed deletion-based diagnostics provide a robust and adaptable framework for understanding and improving recommender systems, bridging the gap between performance optimization and interpretability.

\Urlmuskip=0mu plus 1mu\relax
 \bibliographystyle{elsarticle-num} 
 \bibliography{cas-refs}

\begin{thebibliography}{10}
\expandafter\ifx\csname url\endcsname\relax
  \def\url#1{\texttt{#1}}\fi
\expandafter\ifx\csname urlprefix\endcsname\relax\def\urlprefix{URL }\fi
\expandafter\ifx\csname href\endcsname\relax
  \def\href#1#2{#2} \def\path#1{#1}\fi

\bibitem{schafer.99}
J.~B. Schafer, J.~Konstan, J.~Riedl, Recommender systems in e-commerce, in: Proceedings of the 1st ACM Conference on Electronic Commerce, EC '99, Association for Computing Machinery, New York, NY, USA, 1999, p. 158–166.
\newblock \href {https://doi.org/10.1145/336992.337035} {\path{doi:10.1145/336992.337035}}.

\bibitem{rivas.20}
A.~Rivas, P.~Chamoso, A.~Gonz\'{a}lez-Briones, J.~Pav\'{o}n, J.~M. Corchado, Social network recommender system, a neural network approach, in: Intelligent Data Engineering and Automated Learning – IDEAL 2020: 21st International Conference, Guimaraes, Portugal, November 4–6, 2020, Proceedings, Part II, Springer-Verlag, Berlin, Heidelberg, 2020, p. 213–222.

\bibitem{chang.17}
S.~Chang, Y.~Zhang, J.~Tang, D.~Yin, Y.~Chang, M.~A. Hasegawa-Johnson, T.~S. Huang, Streaming recommender systems, in: Proceedings of the 26th International Conference on World Wide Web, WWW '17, International World Wide Web Conferences Steering Committee, Republic and Canton of Geneva, CHE, 2017, p. 381–389.
\newblock \href {https://doi.org/10.1145/3038912.3052627} {\path{doi:10.1145/3038912.3052627}}.

\bibitem{tintarev.07}
N.~Tintarev, J.~Masthoff, Effective explanations of recommendations: user-centered design, in: Proceedings of the 2007 ACM Conference on Recommender Systems, RecSys '07, Association for Computing Machinery, New York, NY, USA, 2007, p. 153–156.
\newblock \href {https://doi.org/10.1145/1297231.1297259} {\path{doi:10.1145/1297231.1297259}}.

\bibitem{tintarev.072}
N.~Tintarev, Explanations of recommendations, in: Proceedings of the 2007 ACM Conference on Recommender Systems, RecSys '07, Association for Computing Machinery, New York, NY, USA, 2007, p. 203–206.
\newblock \href {https://doi.org/10.1145/1297231.1297275} {\path{doi:10.1145/1297231.1297275}}.

\bibitem{carenini.01}
G.~Carenini, J.~Moore, An empirical study of the influence of user tailoring on evaluative argument effectiveness, in: Proceedings of the Seventeenth International Joint Conference on Artificial Intelligence, IJCAI 2001, Seattle, Washington, USA, August 4-10, 2001, 2001, pp. 1307--1314.

\bibitem{zhou.2023}
G.~Zhou, C.~Huang, X.~Chen, X.~Xu, C.~Wang, L.~Zhu, L.~Yao, Contrastive counterfactual learning for causality-aware interpretable recommender systems, in: Proceedings of the 32nd ACM International Conference on Information and Knowledge Management, CIKM '23, Association for Computing Machinery, New York, NY, USA, 2023, p. 3564–3573.
\newblock \href {https://doi.org/10.1145/3583780.3614823} {\path{doi:10.1145/3583780.3614823}}.

\bibitem{deldjoo.2024}
Y.~Deldjoo, D.~Jannach, A.~Bellogin, A.~Difonzo, D.~Zanzonelli, Fairness in recommender systems: research landscape and future directions, User Modeling and User-Adapted Interaction 34~(1) (2024) 59--108.

\bibitem{chen.2023}
J.~Chen, H.~Dong, X.~Wang, F.~Feng, M.~Wang, X.~He, Bias and debias in recommender system: A survey and future directions, ACM Trans. Inf. Syst. 41~(3) (Feb. 2023).
\newblock \href {https://doi.org/10.1145/3564284} {\path{doi:10.1145/3564284}}.

\bibitem{mansoury.2020}
M.~Mansoury, H.~Abdollahpouri, M.~Pechenizkiy, B.~Mobasher, R.~Burke, Feedback loop and bias amplification in recommender systems, in: Proceedings of the 29th ACM International Conference on Information \& Knowledge Management, CIKM '20, Association for Computing Machinery, New York, NY, USA, 2020, p. 2145–2148.
\newblock \href {https://doi.org/10.1145/3340531.3412152} {\path{doi:10.1145/3340531.3412152}}.

\bibitem{li.2023}
Y.~Li, H.~Chen, S.~Xu, Y.~Ge, J.~Tan, S.~Liu, Y.~Zhang, Fairness in recommendation: Foundations, methods, and applications, ACM Trans. Intell. Syst. Technol. 14~(5) (Oct. 2023).
\newblock \href {https://doi.org/10.1145/3610302} {\path{doi:10.1145/3610302}}.

\bibitem{zhang.2020}
Y.~Zhang, X.~Chen, Explainable recommendation: A survey and new perspectives, Found. Trends Inf. Retr. 14~(1) (2020) 1–101.

\bibitem{vig.09}
J.~Vig, S.~Sen, J.~Riedl, Tagsplanations: explaining recommendations using tags, in: Proceedings of the 14th International Conference on Intelligent User Interfaces, IUI '09, Association for Computing Machinery, New York, NY, USA, 2009, p. 47–56.
\newblock \href {https://doi.org/10.1145/1502650.1502661} {\path{doi:10.1145/1502650.1502661}}.

\bibitem{herlocker.00}
J.~L. Herlocker, J.~A. Konstan, J.~Riedl, Explaining collaborative filtering recommendations, in: Proceedings of the 2000 ACM Conference on Computer Supported Cooperative Work, CSCW '00, Association for Computing Machinery, New York, NY, USA, 2000, p. 241–250.
\newblock \href {https://doi.org/10.1145/358916.358995} {\path{doi:10.1145/358916.358995}}.

\bibitem{sarwar.01}
B.~M. Sarwar, G.~Karypis, J.~A. Konstan, J.~Riedl, Item-based collaborative filtering recommendation algorithms, in: V.~Y. Shen, N.~Saito, M.~R. Lyu, M.~E. Zurko (Eds.), Proceedings of the Tenth International World Wide Web Conference, {WWW} 10, Hong Kong, China, May 1-5, 2001, {ACM}, 2001, pp. 285--295.
\newblock \href {https://doi.org/10.1145/371920.372071} {\path{doi:10.1145/371920.372071}}.

\bibitem{bauman.17}
K.~Bauman, B.~Liu, A.~Tuzhilin, Aspect based recommendations: Recommending items with the most valuable aspects based on user reviews, in: Proceedings of the 23rd ACM SIGKDD International Conference on Knowledge Discovery and Data Mining, KDD '17, Association for Computing Machinery, New York, NY, USA, 2017, p. 717–725.
\newblock \href {https://doi.org/10.1145/3097983.3098170} {\path{doi:10.1145/3097983.3098170}}.

\bibitem{lu.18}
Y.~Lu, R.~Dong, B.~Smyth, Coevolutionary recommendation model: Mutual learning between ratings and reviews, in: Proceedings of the 2018 World Wide Web Conference, WWW '18, International World Wide Web Conferences Steering Committee, Republic and Canton of Geneva, CHE, 2018, p. 773–782.
\newblock \href {https://doi.org/10.1145/3178876.3186158} {\path{doi:10.1145/3178876.3186158}}.

\bibitem{zhang.14}
Y.~Zhang, G.~Lai, M.~Zhang, Y.~Zhang, Y.~Liu, S.~Ma, Explicit factor models for explainable recommendation based on phrase-level sentiment analysis, in: Proceedings of the 37th International ACM SIGIR Conference on Research \& Development in Information Retrieval, SIGIR '14, Association for Computing Machinery, New York, NY, USA, 2014, p. 83–92.
\newblock \href {https://doi.org/10.1145/2600428.2609579} {\path{doi:10.1145/2600428.2609579}}.

\bibitem{abdollahi.17}
B.~Abdollahi, O.~Nasraoui, Using explainability for constrained matrix factorization, in: Proceedings of the Eleventh ACM Conference on Recommender Systems, RecSys '17, Association for Computing Machinery, New York, NY, USA, 2017, p. 79–83.
\newblock \href {https://doi.org/10.1145/3109859.3109913} {\path{doi:10.1145/3109859.3109913}}.

\bibitem{Ribeiro2016}
M.~T. Ribeiro, S.~Singh, C.~Guestrin, "why should i trust you?": Explaining the predictions of any classifier, in: Proceedings of the 22nd ACM SIGKDD International Conference on Knowledge Discovery and Data Mining, 2016, pp. 1135--1144.

\bibitem{Tohidi2024}
N.~Tohidi, M.~Beheshti, Enhanced explanations in recommendation systems, in: 2024 IEEE International Symposium on Systems Engineering (ISSE), 2024, pp. 1--5.

\bibitem{Koh2017}
P.~W. Koh, P.~Liang, Understanding black-box predictions via influence functions, in: Proceedings of the 34th International Conference on Machine Learning, PMLR, 2017, pp. 1885--1894.

\bibitem{He.2017}
X.~He, L.~Liao, H.~Zhang, L.~Nie, X.~Hu, T.-S. Chua, Neural collaborative filtering, in: Proceedings of the 26th International Conference on World Wide Web, WWW '17, International World Wide Web Conferences Steering Committee, Republic and Canton of Geneva, CHE, 2017, p. 173–182.

\bibitem{Ponnusamy.23}
C.~Ponnusamy, W.-K. Wong, A.~Raja, O.~Khalaf, A.~Kiran, J.~Babu, Health recommendation system using deep learning-based collaborative filtering, Heliyon 9 (2023) e22844.
\newblock \href {https://doi.org/10.1016/j.heliyon.2023.e22844} {\path{doi:10.1016/j.heliyon.2023.e22844}}.

\bibitem{Mulyana_Rumaisa_2024}
H.~L. Mulyana, F.~Rumaisa, Course learning recommendation system using neural collaborative filtering, Brilliance: Research of Artificial Intelligence 4~(2) (2024) 517--524.
\newblock \href {https://doi.org/10.47709/brilliance.v4i2.4699} {\path{doi:10.47709/brilliance.v4i2.4699}}.

\bibitem{marzuki.2024}
I.~Marzuki, M.~Hariadi, R.~Rachmadi, Y.~Arif, Neural collaborative filtering for improved tourism destination recommendation, in: 2024 8th International Conference on Information Technology, Information Systems and Electrical Engineering (ICITISEE), 2024, pp. 481--486.
\newblock \href {https://doi.org/10.1109/icitisee63424.2024.10730542} {\path{doi:10.1109/icitisee63424.2024.10730542}}.

\bibitem{wei.2025}
C.~Wei, Tourist attraction image recognition and intelligent recommendation based on deep learning, Journal of computational methods in sciences and engineering (2025 FEB 7 2025).
\newblock \href {https://doi.org/10.1177/14727978251318805} {\path{doi:10.1177/14727978251318805}}.

\bibitem{venkatesh.2024}
V.~C, H.~Oberoi, A.~Goyal, N.~Sikka, Re-recsys: An end-to-end system for recommending properties in real-estate domain, in: Proceedings of the 7th Joint International Conference on Data Science; Management of Data (11th ACM IKDD CODS and 29th COMAD), CODS-COMAD 2024, ACM, 2024, p. 558–562.
\newblock \href {https://doi.org/10.1145/3632410.3632487} {\path{doi:10.1145/3632410.3632487}}.

\bibitem{koren2009matrix}
Y.~Koren, R.~Bell, C.~Volinsky, Matrix factorization techniques for recommender systems, Computer 42~(8) (2009) 30--37.

\bibitem{abdollahpouri2017controlling}
H.~Abdollahpouri, R.~Burke, B.~Mobasher, Controlling popularity bias in learning-to-rank recommendation, in: Proceedings of the 11th ACM Conference on Recommender Systems, ACM, 2017, pp. 42--46.

\bibitem{steck2018calibrated}
H.~Steck, Calibrated recommendations, Proceedings of the 12th ACM Conference on Recommender Systems 2018 (2018) 154--162.

\bibitem{jadon2024}
A.~Jadon, A.~Patil, \href{arxiv.org/html/2312.16015v2}{A comprehensive survey of evaluation techniques for recommendation systems} (2024).
\newblock \href {http://arxiv.org/abs/2312.16015} {\path{arXiv:2312.16015}}.
\newline\urlprefix\url{arxiv.org/html/2312.16015v2}

\bibitem{zangerle.2022}
E.~Zangerle, C.~Bauer, Evaluating recommender systems: Survey and framework, ACM Computing Surveys 55~(8) (2022) 1--38.

\bibitem{Harper.2015}
F.~M. Harper, J.~A. Konstan, The movielens datasets: History and context, ACM Trans. Interact. Intell. Syst. 5~(4) (Dec. 2015).
\newblock \href {https://doi.org/10.1145/2827872} {\path{doi:10.1145/2827872}}.

\bibitem{rendle.2020}
S.~Rendle, W.~Krichene, L.~Zhang, J.~Anderson, \href{arxiv.org/abs/2005.09683}{Neural collaborative filtering vs. matrix factorization revisited} (2020).
\newblock \href {http://arxiv.org/abs/2005.09683} {\path{arXiv:2005.09683}}.
\newline\urlprefix\url{arxiv.org/abs/2005.09683}

\bibitem{Kammoun.2022}
A.~Kammoun, R.~Slama, H.~Tabia, T.~Ouni, M.~Abid, Generative adversarial networks for face generation: A survey, ACM Computing Surveys (Mar. 2022).
\newblock \href {https://doi.org/10.1145/1122445.1122456} {\path{doi:10.1145/1122445.1122456}}.

\bibitem{fan.2024}
H.~Fan, M.~Zhu, Y.~Hu, H.~Feng, Z.~He, H.~Liu, Q.~Liu, \href{arxiv.org/abs/2409.16182}{Tim4rec: An efficient sequential recommendation model based on time-aware structured state space duality model} (2024).
\newblock \href {http://arxiv.org/abs/2409.16182} {\path{arXiv:2409.16182}}.
\newline\urlprefix\url{arxiv.org/abs/2409.16182}

\bibitem{chen.2024}
Y.~Chen, J.~Tan, A.~Zhang, Z.~Yang, L.~Sheng, E.~Zhang, X.~Wang, T.-S. Chua, \href{arxiv.org/abs/2406.09215}{On softmax direct preference optimization for recommendation} (2024).
\newblock \href {http://arxiv.org/abs/2406.09215} {\path{arXiv:2406.09215}}.
\newline\urlprefix\url{arxiv.org/abs/2406.09215}

\bibitem{mcauley.2013}
J.~McAuley, J.~Leskovec, Hidden factors and hidden topics: understanding rating dimensions with review text, in: Proceedings of the 7th ACM Conference on Recommender Systems, RecSys '13, Association for Computing Machinery, New York, NY, USA, 2013, p. 165–172.
\newblock \href {https://doi.org/10.1145/2507157.2507163} {\path{doi:10.1145/2507157.2507163}}.

\end{thebibliography}





\end{document}